## soft matter
## Novel core-shell structures for colloids made from surfactants and polymers

**J.–F. Berret and P. Hervé** *(CNRS - Cranbury Research Center Rhodia Inc., Cranbury, USA)*

**I. Grillo** *(ILL)*

We report on the formation of colloidal complexes resulting from the electrostatic self-assembly of polyelectrolyte-neutral block copolymers and oppositely charged surfactants. The copolymers investigated are asymmetric and characterised by a large neutral block. Using light, neutron and x-ray scattering experiments, we have shown that the colloidal complexes exhibit a core-shell microstructure. The core is described as a dense and disordered micro-phase of micelles connected by the polyelectrolyte blocks, whereas the shell is a diffuse brush made from the neutral chains. Colloidal complexes with core-shell structures resemble the well-known amphiphilic block copolymer micelles. The self-assembly mechanism is however different here. It is based on the complexation between opposite charges. This mechanism has recently attracted much interest because it allows the association of components of different nature, such as organic and inorganic or synthetic and biological.

**During the last decade,** there has been a widespread interest for the design and synthesis of new polymer-based colloidal particles of increased stability in aqueous media. Among these particles, colloidal complexes have emerged as a new type of microstructure, with potential applications in home and personal care, catalysis and drug delivery. Colloidal complexes result from a self-assembly mechanism between polyelectrolyte-neutral co-polymers and oppositely charged species. The first macromolecules or macro-ions having been complexed with these polymers were synthetic and biological macromolecules [1,2], proteins [3] and surfactant micelles [4]. In this report, we re-examine the case of surfactant micelles using small-angle neutron scattering. Two systems are compared. The first (1) is a dilute aqueous solution containing a polyelectrolyte (poly(sodium-acrylate) and a surfactant (dodecyl trimethyl/ammonium bromide (DTAB)) of opposite charge. The total concentration of active matter is a few percent (by weight) and the charge ratio, Z is 1. Z is the ratio of the number of surfactant molecules to charged monomer. The second system (2) is an aqueous solution at the same concentration and same charge ratio, where the polyelectrolyte is replaced by a diblock co-polymer. The first block of the co-polymer is the same charged chain as in system (1) and the second block is neutral and water-soluble [5]. Details of the structure and molecular weights are given in the captions.

In system (1), the polymer-surfactant solution undergoes a phase separation resulting in two well separated and stable thermodynamic phases. The lower phase is a white precipitate, whereas the upper phase is fluid and transparent. Figure 1 shows the x-ray scattering intensity of the precipitate. The 10 Bragg reflections observed between 0.1 Å$^{-1}$ and 0.3 Å$^{-1}$ attest to the existence of a long-range cubic order, associated with the space group *Pm3n* [6]. The cubic symmetry suggests that the surfactants are organized in spherical micelles, these micelles being at the nodes of the cubic lattice (figure 2). In this arrangement, the polyelectrolytes are wrapped around the micelles and can connect several of them.

When the former homopolyelectrolyte chain is now part of a diblock co-polymer, the phase separation described previously ceases to occur [6,7]. For system (2), the mixed solutions appear homogeneous and transparent. Figure 3 shows the scattering cross-section measured by light and small-angle neutron scattering on such solutions. d$\sigma$/d$\Omega$(q) is dominated by the two features : a strong forward scattering (at low wave-vectors) and a structure peak at high wave-vectors. In order to interpret the spectra of figure 3 quantitatively, we have assumed for the surfactant-polymer com-plexes a core-shell structure similar to that shown in figure 4 [6-9]. In this model, the core is a dense and disordered assembly of surfactant micelles connected by the polyelectrolyte blocks. The corona is a diffuse shell of the neutral chains which is detected at much lower wave-vectors. Dynamic light scattering experiments actually show rather monodisperse hydrodynamic sizes around 50 nm (radius), which we attribute to the overall dimension of the colloids. The continuous line in figure 3 has been calculated by Monte Carlo simulations using a hard sphere interaction potential between micelles [9], and an average aggregation number of 105 micelles. The agreement between experimental data and the model is excellent, especially at high q where the structure factor is correctly reproduced.

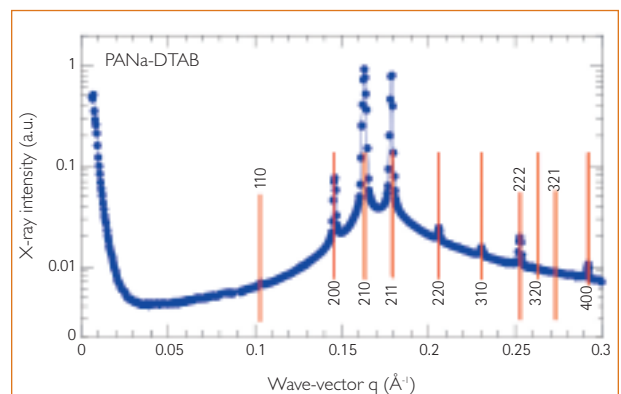

**Figure 1:** X-ray scattering intensity of the precipitate obtained by mixing poly(sodium acrylate) and dodecyltrimethylammonium bromide solutions at total concentration c = 2 wt. % and charge ratio Z = 1. The molecular weight of the polyelectrolyte is 30 000 g·mol$^{-1}$, corresponding to 420 monomers. The succession of Bragg peaks is consistent with a cubic structure of symmetry *Pm3n*. The lattice parameter of the elementary cell is 8.61 nm [6].

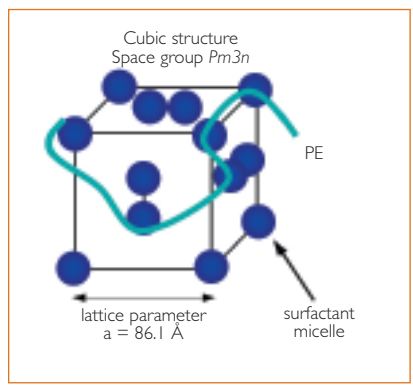

**Figure 2:** Schematic representation of the cubic structure associated with the space group of symmetry *Pm3n*. The spheres are the surfactant micelles (radius 2 nm) and the polyelectrolytes are assumed to wrap around them. For *Pm3n* cubic structure, each face of the lattice cell has two surfactant micelles and the micellar volume fraction is 0.524.

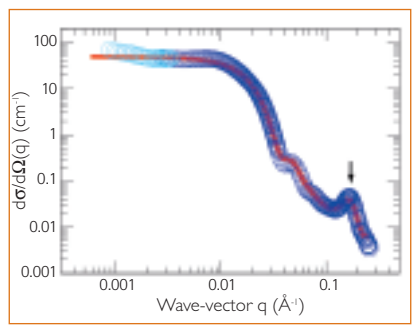

**Figure 3:** Light and neutron scattering intensities of solutions obtained by mixing poly(sodium acrylate)-b-poly(acrylamide) and dodecyltrimethylammonium bromide solutions at total concentration c = 1 wt.% and at charge ratio Z = 1 [7]. The neutron scattering data were collected on the D22 spectrometer at the ILL. The molecular weights for both blocks are 5 000 g·mol$^{-1}$ for the charged segment and 30 000 g·mol$^{-1}$ for the neutral one. For this solution, the solvent is D$_2$O. An arrow indicates the position of a structure peak at 0.16 Å$^{-1}$. The peak corresponds to an interaction distance of ~ 4 nm, *i.e.* twice the radius of a micelle. Continuous line : calculation of the form factor of the mixed aggregate represented in figure 4 [9]. Here, we assume a distribution in aggregation numbers, around an average value of 105 and with a standard deviation of 52.

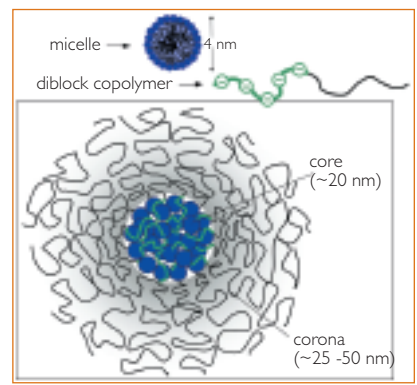

**Figure 4:** Representation of a colloidal complex formed through the association of oppositely charged block copolymers and surfactant micelles. Depending on the molecular weights of the copolymers, the radius of the core ranges from 10 to 20 nm and the corona thickness between 25 and 50 nm [6].

Co-polymers of different molecular weights and of different nature have been studied in order to demonstrate the universality of the electrostatic self-assembly process using surfactants. Eight different polymers have been investigated and hierarchical structures similar to that of figure 4 were observed [6]. We have found that the size of the colloids depends predominantly on the degree of polymerisation of the charged block, and not on that of the neutral chain. We suggest that the mechanism of formation of the colloidal complexes is identical to that of the phase transition encountered in the polyelectrolyte-surfactant mixtures. With the co-polymers, the growth processes are arrested at a stage that is controlled by the neutral blocks. The polymer brush surrounding the core not only prevents the macroscopic phase separation. It also stabilizes the aggregate.

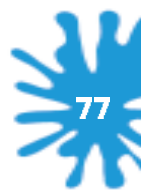

77